\documentclass[twocolumn,superscriptaddress,showpacs,preprintnumbers,amsmath,amssymb]{revtex4-1}

\usepackage{graphicx}
\usepackage{dcolumn}
\usepackage{bm}

\begin{document}
 \preprint{APS/123-QED}
 
\title{Non-equilibrium steady states of stochastic processes with intermittent resetting}

\author{Stephan Eule}
\affiliation{Max Planck Institute for Dynamics and Self-Organization, Am Fa{\ss}berg 17, 37077 G\"ottingen, Germany}

\author{Jakob J.~Metzger}
\altaffiliation[Present address: ]{The Rockefeller University, 1230 York Avenue, 10065 New York, NY, USA}

\affiliation{Max Planck Institute for Dynamics and Self-Organization, Am Fa{\ss}berg 17, 37077 G\"ottingen, Germany}
\affiliation{Institute for Nonlinear Dynamics, Department of Physics, University of G\"ottingen, Germany}

\date{\today}

\begin{abstract}
Stochastic processes that are randomly reset to an initial condition serve as a showcase to investigate non-equilibrium steady states. However, all existing results have been restricted to the special case of memoryless resetting protocols. Here, we obtain the general solution for the distribution of processes in which waiting times between reset events are drawn from an arbitrary distribution. This allows for the investigation of a broader class of much more realistic processes. As an example, our results are applied to the analysis of the efficiency of constrained random search processes.
\end{abstract}

\maketitle

Suppose that you are working on a difficult problem. 
While trying to find a solution, you may get the impression that you are stuck or got on the wrong track.
A natural strategy in such a situation is 
to reset from time to time and start over.
This behavior can be modeled by a stochastic exploration process, which is interrupted by a random resetting to
the initial condition. Over the last years a special case of such processes, a diffusion process interrupted at constant rate
by reset events, has attracted considerable attention, because it represents a particularly simple and analytically approachable example of a non-equilibrium steady state (NESS) \cite {Evans, Evans2, Evans3, Pal, Majum}.

The investigation of processes with random resetting is also one of natural interest to the
study of first passage times, e.g.~in the catalysis time of chemical reactions \cite{Reuveni}, in kinetic proofreading \cite{Hopfield}, and in areas where search optimality is crucial \cite{Shlesinger}. Furthermore, processes with reset are studied
in population dynamics, where resets are interpreted as catastrophic events corresponding to the extinction of the population,
followed by a resurgence \cite{catastrophe}.

All of the previous works, including the extensions to L\'evy Flights \cite{Kusmierz} and fluctuating interfaces with stochastic resetting \cite{Gupta}, assume the special case that resets occur at a constant rate $\gamma$, i.e.~the case in which the waiting times between the resets are exponentially distributed.
In this case the resetting procedure is memoryless, which has a straight-forward but interesting implication:
The NESS of the process is equivalent to the distribution of end-points of an ensemble of trajectories of the same process without resetting that are trapped at a constant rate $\gamma$. Consequently, if the propagator of the process without resetting is known, the NESS
    can be directly obtained by averaging the propagator over the exponential distribution of inter-reset times \cite{Evans}.  
  \begin{figure}[ht!]
    \includegraphics[width=1.\columnwidth]{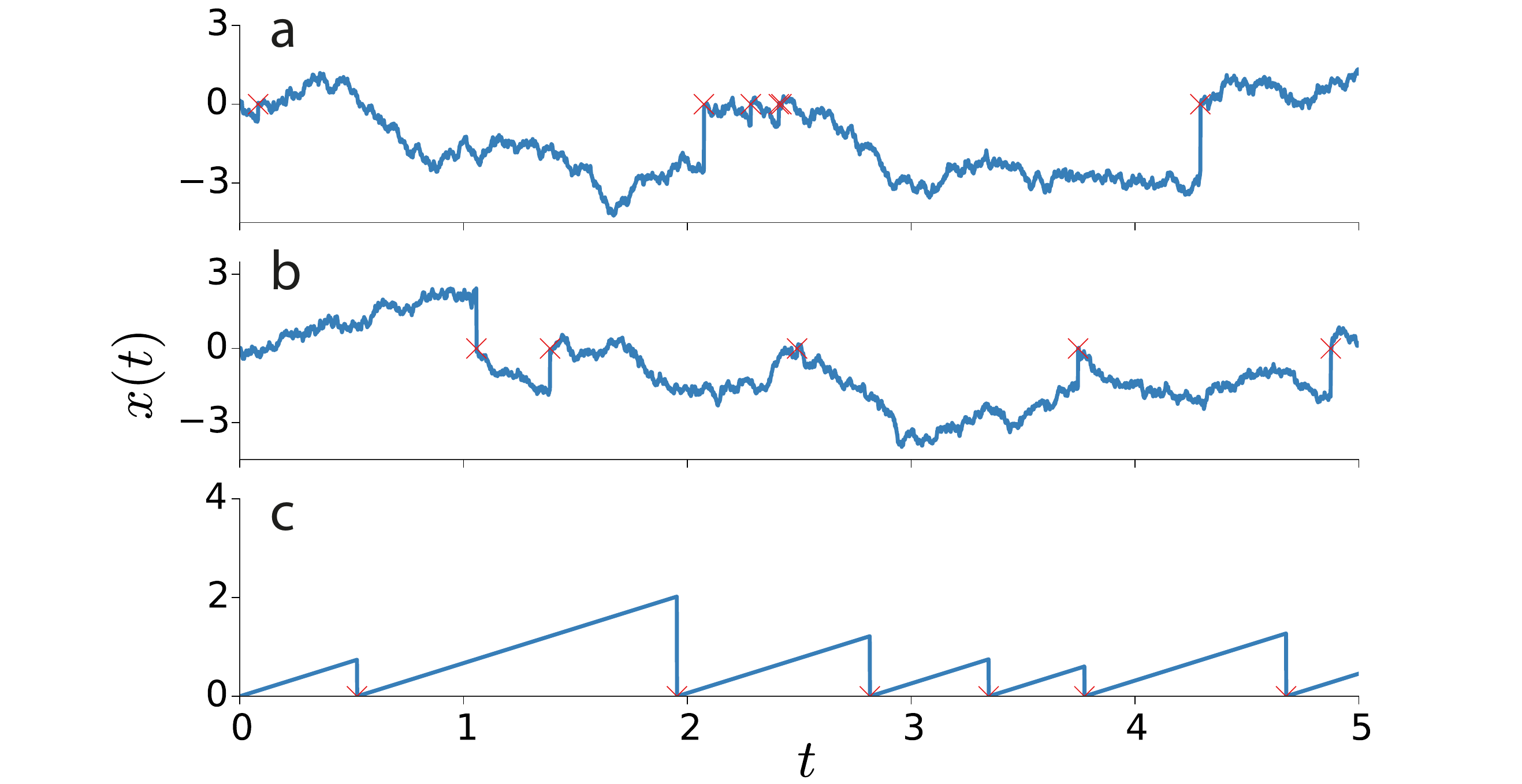}
  \caption{Examples of different processes $x(t)$ with random resetting (marked with red crosses). All cases share the same mean duration between resets, $\langle\tau\rangle=1$, but their distributions differ. (a) Diffusion with intermittent (bursty) resetting. A resetting event is more likely to occur when another one has happened recently. Distribution of resetting times is heavy tailed. (b) Diffusion with resetting times that are comparatively regular (distribution of resetting times peaked around mean). (c) Stochastic resetting with deterministic, linear motion between resetting events. }
  \label{fig1}
\end{figure} 
However, the limitation to constant rate resetting severely restricts the applicability to memoryless processes.
Even the simple example of a process in which multiple identical steps have to be completed for a reset, which gives rise to Gamma-distributed waiting times, cannot be described within the current framework. Such processes occur e.g.~in chemical resetting due to multistep dissociation reactions \cite{Helferrich}. Neither can the important case of a rate that depends on the time $\tau$ elapsed since the last reset event be captured anymore. Processes of this form can be described using time-dependent rates $\tilde{\gamma}(\tau)=(\alpha\gamma) (\tau\gamma)^{\alpha-1}$, which leads to Weibull distributed waiting times. Here the probability for a event decreases (increases) with $\tau$ for $\alpha<1$ ($\alpha>1$), thus giving
rise to more intermittent (regular) reset sequences, as compared to the constant rate case ($\alpha=1$). The Weibull distribution is widely applied
when the conditions for strict history-independence are violated \cite{Weibull}. 


In this Letter, we present the theory of stochastic resetting with arbitrary waiting time distributions $\psi(\tau)$,  (cf.~Fig.~\ref{fig1}),
which includes the aforementioned examples as special cases.
While for stochastic processes that reset at a constant rate, the master equation can be immediately formulated, here
we first have to derive the governing equation for arbitrary resetting time distributions. The corresponding master equation Eq.~(\ref{main}) is non-local 
in time and is one of our main results. Despite the fact that our theory has to include integrals over the complete history of the process, we can explicitly calculate stationary distributions for different waiting time distributions, $\psi(\tau)$, and quantify the temporal relaxation of the moments towards this NESS.
We find that the history-dependent processes with a general resetting time distribution have a rich structure and exhibit many new properties that are not present in the case of resetting with a constant rate. Furthermore, we demonstrate that the success of a search depends on the full waiting time distribution and not only on the characteristic resetting time scale as one could expect from the memoryless process. 
A possible application of our theory is the efficiency of random searchers that are confined by their need to 
regularly return to a home location.
We find that, under this constraint, the search success can be optimized 
by adapting the distribution of waiting times between the returns. 


Consider the motion of a particle that, between reset events, 
is described by the stochastic differential equation
\begin{equation}\label{SDE}
{\dot x}(t)= F(x)+\xi(t)\, ,
\end{equation}
where $\xi(t)$ is Gaussian white noise with correlation $\langle  \xi(t)\xi(t')\rangle =2D\delta(t-t')$. Furthermore, let
the waiting time distribution between the resets be $\psi(\tau)$, after which the particle is reset to the location $x_r$.
The probability $\nu_n(t)$ that the $n$'th reset event happens at time $t$ satisfies the renewal equation
$
\nu_n(t)=\int_0^t \psi(t-t') \nu_{n-1}(t') dt'\, ,
$
stating that the probability to have the $n$'th event at time $t$ is given by the probability that the $n-1$'th event occurred at time $t'$ 
and that the next event occurs after time $t-t'$ \cite{Cox}. We assume that the process starts with
a reset event. Then, the probability that an event occurs at time $t$ is 
$\nu(t)=\delta(t)+\sum_{n=1}^{\infty} \nu_n(t)$, where the $\delta$-distribution accounts for the initial condition.
The probability $p(x, t)$ to find the particle at location $x$ at time $t$ is
\begin{equation}\label{peq}
p(x, t)=\int_0^t w(t-t') p(x, x_r; t-t') \nu(t') dt' \, ,
\end{equation}
where $p(x, x_r;t-t')$ is the transition amplitude of the process defined by Eq.~(\ref{SDE}). 
The distribution $w(t-t')$ gives the probability that no resetting event
occurs between $t'$ and $t$ and is related to the waiting time distribution $\psi(\tau)$ according to
$w(\tau)=1-\int_0^\tau \psi(t) dt$.
To derive an evolution equation for $p(x,t)$, we switch to the Laplace domain ${\hat p}(x,s) = \int_0^\infty dt\,e^{-st} p(x,t)$.
Using the operator representation $p(x, x_r; t-t')={e^{(t-t')\mathcal{L}(x)}\delta(x-x_r)}$ \cite{Risken} 
and applying the convolution and shifting theorem, the Laplace transform of Eq.~(\ref{peq}) reads
\begin{equation}\label{peqlap}
{\hat p(x, s)}= {\hat w}(s-\mathcal{L}(x)) \delta(x-x_r){\hat \nu}(s).
\end{equation}
Noting that the Laplace transform of $\nu(t)$ is 
\begin{equation}\label{nu}
{\hat \nu}(s)=1+{\hat\psi}(s)/(1-{\hat\psi}(s))\,,
\end{equation} 
we can use Eq.~(\ref{peqlap}) and Eq.~(\ref{nu}) to obtain
\begin{align}\label{Laplacefull}
s\, {\hat p}(x,s)-p_0(x) =&\mathcal{L}(x)\, {\hat p}(x,s)\\
&+\frac{{\hat \phi}(s)}{s}\delta(x-x_r)  -{\hat \phi}(s-\mathcal{L}(x)){\hat p}(x, s)\, , \nonumber
\end{align}
where the time evolution kernel ${\hat \phi}(s)$ is given by
$
{\hat \phi}(s)=s {\hat \psi}(s)/(1-{{\hat \psi} }(s))
$
\cite{Friedrich} and $p_0(x)=\delta(x-x_r)$ is the initial condition.
Finally, Laplace inversion of Eq.~(\ref{Laplacefull}) leads to the generalized master equation
\begin{align}\label{main}
\frac{\partial}{\partial t} p(x, t) &=\mathcal{L}(x)\, p(x, t) \\
& +\int_0^t\phi(t-t')\left[ \delta(x-x_r)- e^{(t-t'){\mathcal L}(x)} p(x, t') \right] dt'\, . \nonumber
\end{align}
The first term on the r.h.s accounts for the dynamics between the resetting events, the second term is the source term and
describes the resets to $x_r$. The third term is a sink term; it describes the density of particles that have propagated to
$x$ at time $t$ during the time $t-t'$ under the influence of the dynamics Eq.~(\ref{SDE}) and are then subject to resetting. 
The integral of the time-evolution kernel corresponds to the time-dependent density of reset events, i.e.~$\nu(t)=\int_0^t\phi(\tau) d\tau$.
The special case of resetting at a constant rate $\gamma$ is recovered by choosing an exponential waiting time distribution 
$\psi(\tau)=\gamma\,e^{-\gamma \tau}$ that leads to $\phi(\tau)=\gamma\delta(\tau)$. Inserting this kernel
into Eq.~(\ref{main}) leads to the Markovian master equation studied in \cite{Evans}. 
The solution of Eq.~(\ref{main}) in Laplace space is
\begin{align}\label{Laplacesol}
{\hat p}(x,s)=& \frac{p_0(x)}{s-\mathcal{L}(x)+{\hat \phi}(s-\mathcal{L}(x))}\\ 
& +\frac{{\hat \phi}(s) \delta(x-x_{r})}{s \left[ s-\mathcal{L}(x)+{\hat \phi}(s-\mathcal{L}(x)) \right]}\, . \nonumber
\end{align}
While the Laplace inversion of Eq.~(\ref{Laplacesol}) can in general only be carried out by numerical means, we will show that the time-dependent moments of the distribution can be calculated analytically. Remarkably, even the full stationary solution $p_\mathrm{st}(x)=\lim_{t \to \infty }p(x,t)$ can be given by employing the Tauberian theorems \cite{Feller}, which allow to consider instead the limit $\lim_{s \to 0}{\hat p}(x,s)$ in Laplace space.

\begin{figure}[htb]
  \centering
    \includegraphics[width=1.\columnwidth]{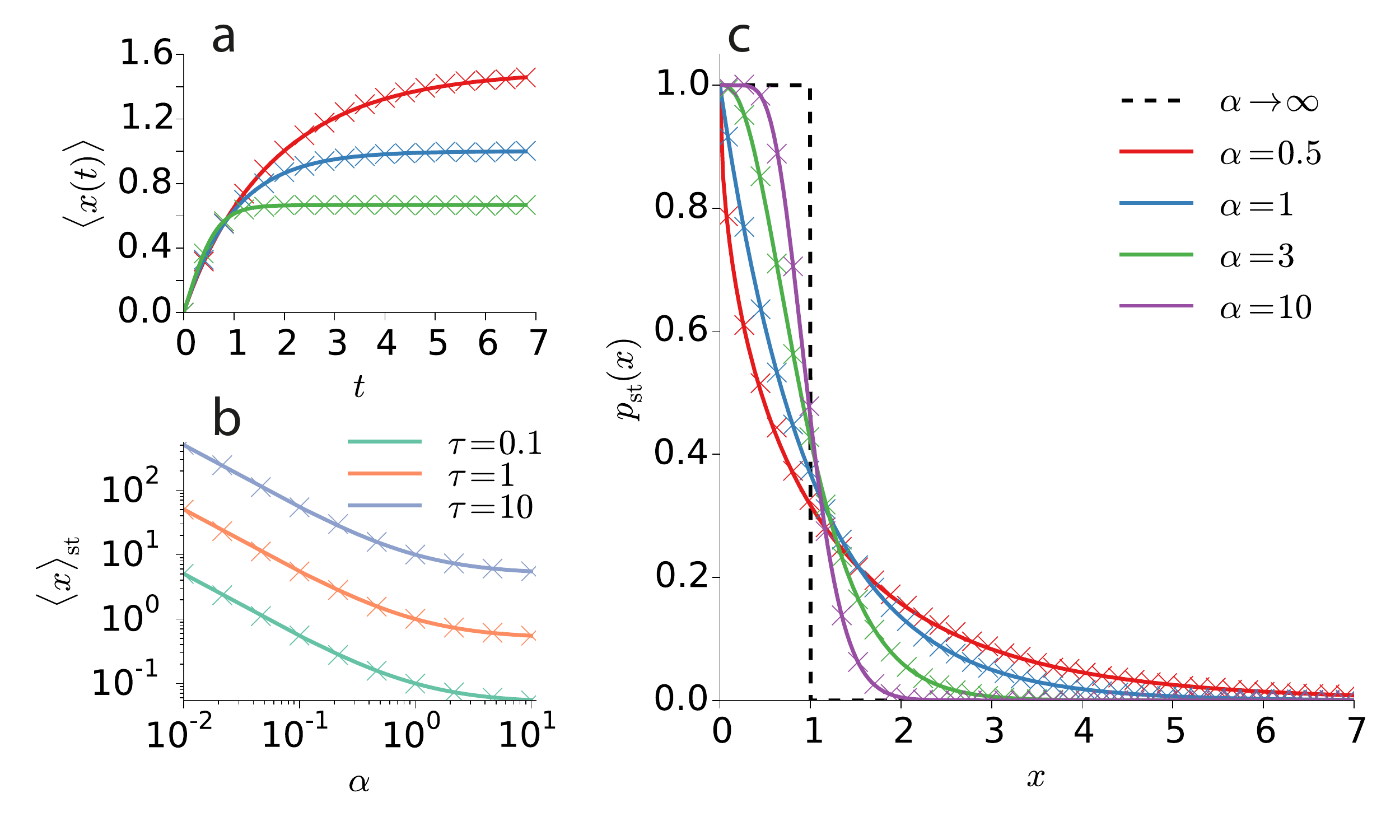}
  \caption{Analytical (lines) and numerical (crosses) results for the case of advection, with resetting times drawn from a Gamma distribution with different intermittency parameters $\alpha$. (a) Time evolution of the mean position in the case of advection (legend see panel c). More intermittent resetting (smaller $\alpha$) leads to an increased stationary mean $\langle x \rangle _\mathrm{st}$, although the mean time between resets is kept constant at $\langle\tau\rangle=1$. (b) Stationary value $\langle x \rangle _\mathrm{st}$ for a range of intermittency parameters $\alpha$ and different mean resetting times $\langle\tau\rangle$. (c) Full stationary distribution for different $\alpha$ and $\langle\tau\rangle=1$. For very large $\alpha$ the distribution tends to a step function.}
  \label{fig:advection}
\end{figure}


Before we come to the diffusion processes with resetting, let us first exemplify Eq.~(\ref{main}) using the example of constant advection with velocity $v$ between the reset events (Fig.~\ref{fig1}c). In particular, we set $F(x)=v$ and $D=0$ in Eq.~(\ref{SDE}) and
consider the Liouvillian $\mathcal{L}(x)= -v\partial_x$. Without loss of generality let us from now on assume the reset position to be $x_r=0$. 
Using Eq.~(\ref{main}), we obtain for the time evolution of the first moment in Laplace space
\begin{equation} \label{x_mean}
\langle  {\hat x}(s) \rangle = \frac{v}{s(s+ {\hat \phi}(s))}\left( 1+\frac{d}{ds}{\hat \phi}(s) \right)\, .
\end{equation}
Clearly, the mean of the process depends on the precise form of the waiting time distribution through ${\hat \phi}(s)$ even in the stationary limit.
For concreteness, let us consider the waiting times to be Gamma-distributed with rate $\gamma$ and shape parameter $\alpha$, i.e.~
$\psi(\tau; \alpha,\gamma)=\tau^{\alpha-1}e^{-\gamma\tau}\gamma^\alpha/ \Gamma(\alpha)$. The mean of this distribution is
$\langle \tau\rangle =\alpha/\gamma$ and for $\alpha=1$ it includes the exponential distribution. The shape parameter $\alpha$ regulates how
intermittently the resetting events occur. For $\alpha<1$ these events occur more intermittently the smaller $\alpha$ gets,
whereas for $\alpha\to \infty$ the resetting events occur regularly with period $\langle \tau \rangle $.
In Fig.~\ref{fig:advection}a we compare $\langle x(t)\rangle $ for Gamma-distributed waiting times with the same mean waiting time but with different rate and shape parameters, calculated using Eq.~\eqref{x_mean}. Interestingly, the asymptotic value of $\langle x(t) \rangle$ depends on the shape of the distribution, even when the mean time between resets, $\langle\tau\rangle$, is kept constant.
We can also calculate the asymptotic, steady-state value of the mean by expanding Eq.~\eqref{x_mean} in $s$, keeping only the lowest order, and then performing the Laplace inversion. This leads to $\langle x \rangle _{st}=\frac{v \langle  \tau \rangle }{2}\frac{1+\alpha}{\alpha}$, which only for exponentially distributed waiting times is equal to the typical distance $x_{s}=v\langle \tau \rangle$  that a particle moves between resets (Fig.~\ref{fig:advection}b). 
The stationary distribution can be calculated using the same approach. Inserting the Liouvillian $\mathcal{L}(x)= -v\partial_x$ in Fourier space, $\mathcal{L}(k)=ivk$, in Eq.~\eqref{Laplacesol}, and substituting the appropriate $\hat \phi(s)$, 
the resulting expression can be considered in the limit $s \to 0 $. For example, in the case of Gamma-distributed resetting times, the resulting expression in Fourier space is $p_{st}(k)=i \gamma  (1-(1-i k /\gamma )^{-\alpha})/(\alpha  k)$, which can easily be Fourier inverted. Several examples are given in Fig.~\ref{fig:advection}c. The stationary distribution
exhibits a transition
from a broad-tailed distribution for $\alpha <1$, where rare large excursions dominate the process, to a uniform distribution in the limit $\alpha\to \infty$,
where the process is reset deterministically after time $\langle \tau\rangle $. 


Let us now focus on the case of free diffusion between the resets, i.e.~$F(x)=0$, with $\mathcal{L}(x)=D \partial^2_x$. In this case, the process is symmetric around the reset point and the mean is simply $x_r$ (set to zero for simplicity). Interestingly, calculating the second moment of the process, $\langle x^2(t)\rangle$ from Eq.~\eqref{main}, leads to exactly the same equation as the mean of the advection process, but with $v$ replaced by $2D$. Thus, considering the time evolution of the variance and its final state leads to exactly the same curves as shown for the mean of the advection in Fig.~\ref{fig:advection}a and b, as we have also confirmed numerically. The stationary distribution $p_\text{st}(x)$
can again be calculated 
by considering Eq.~\eqref{Laplacesol} in the limit of small $s$. For the Gamma-distribution, one obtains after Laplace inversion
\begin{align}\label{gammasol}
&p_{st}(x)= \tfrac{1}{\sqrt{\pi d \gamma}}\  _1F_2(-\tfrac{1}{2};\tfrac{1}{2},\tfrac{1}{2}-\alpha ;\tfrac{x^2}{4 d \gamma}) - \frac{|x|}{2d a \gamma}\\
&+ \frac{|x|^{1+2\alpha}\Gamma(-2\alpha-1)}{(d \gamma)^{\alpha+1}\pi \alpha} \ _1F_2(\alpha ;\alpha +1,\alpha
   +\tfrac{3}{2};\tfrac{x^2}{4 d \gamma }) \sin(\pi \alpha) \nonumber
\end{align}
where $_{q}F_{r}$ denotes the generalized hypergeometric function. The general solution Eq.~\eqref{gammasol} reduces to particularly simple expressions for integer values of $\alpha$, e.g.~$\exp(-|x|/\sqrt{d\gamma})(|x|+3\sqrt{d\gamma})/(8d\gamma)$ for $\alpha{=}2$.
\begin{figure}[htb]
  \centering
    \includegraphics[width=1.\columnwidth]{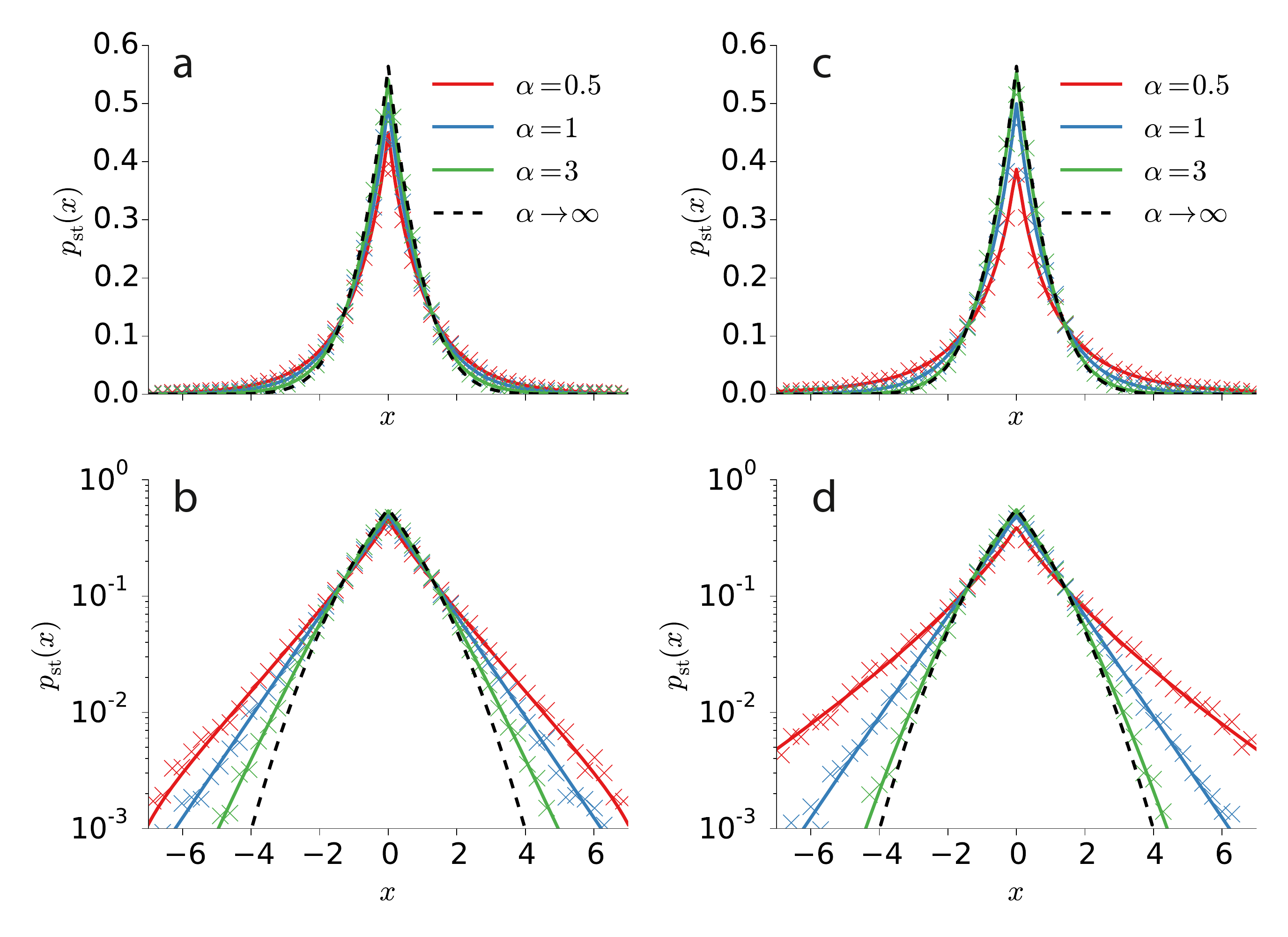}
  \caption{Stationary distributions for diffusion between reset events for the (a,b) Gamma and (c,d) Weibull distributions (lower panels: log-scale). The mean resetting time is fixed to $\langle\tau\rangle = 1$. Numerical simulations (crosses) perfectly match the analytic solutions (solid lines). Smaller shape parameter $\alpha$ corresponds to more intermittent resetting and leads to broader tailed stationary distributions. Both cases lead to the same limiting distribution for large $\alpha$ (dashed line) corresponding to regular resetting at $\langle\tau\rangle$, see Eq.(\ref{det}).}
  \label{fig:pics_FIG_SD}
\end{figure}
\begin{figure}[htb]
  \centering
    \includegraphics[width=1.\columnwidth]{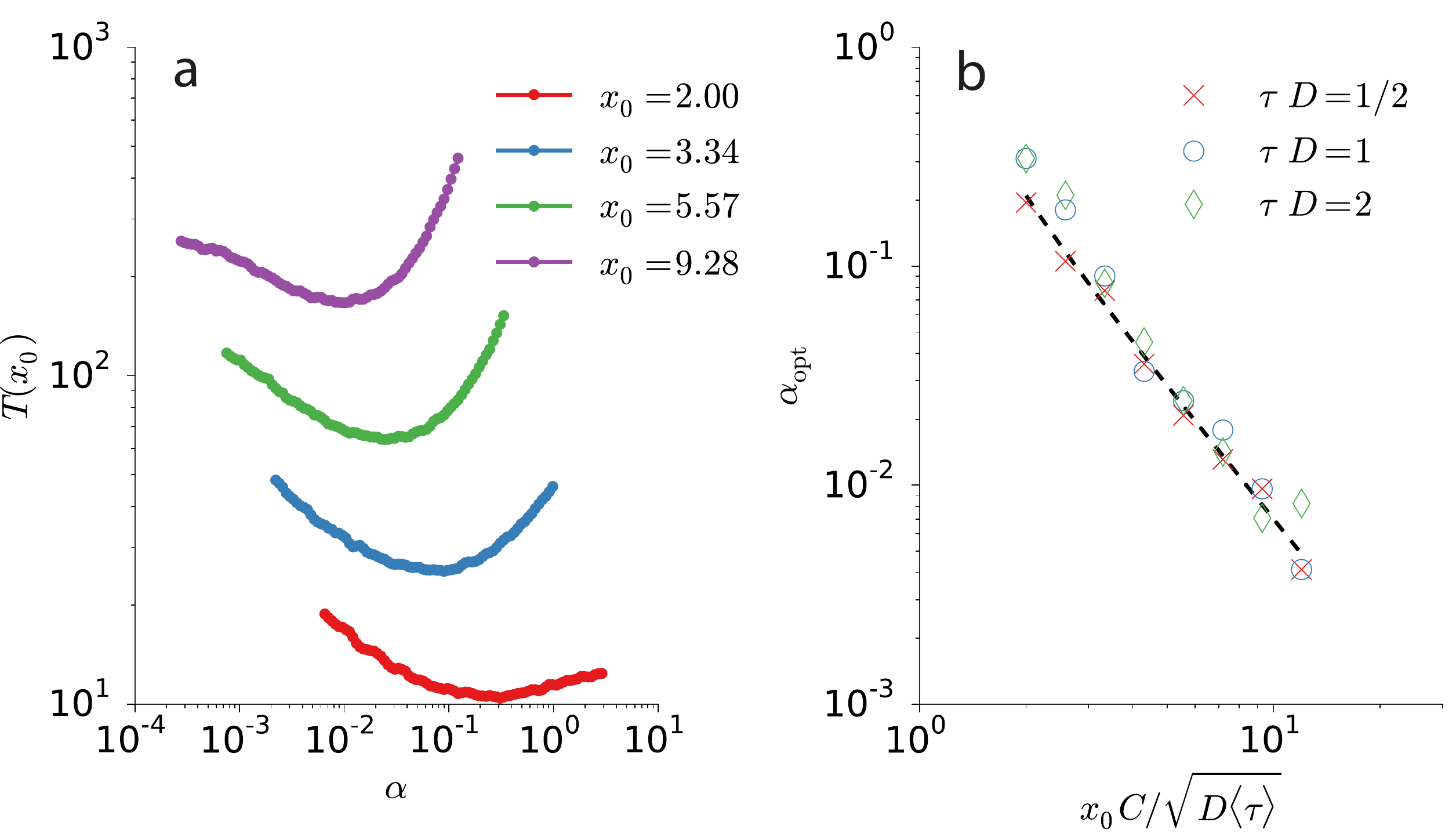}
  \caption{(a) Numerically determined mean first passage time $T(x_0)$ to a distance $x_0$ for constant mean resetting time $\langle\tau\rangle$ and varying intermittency $\alpha$ for the case of Gamma-distributed reset times. A clear minimum of the mean first passage time can be observed at a specific optimal $\alpha_{opt}$. (b) Optimal values of $\alpha$ for different target distances $x_0$ and different parameters of the resetting distribution and the diffusion constant. The curves collapse to the predicted expression, Eq.~\eqref{alphaopt}.}
  \label{fig4}
\end{figure}
In Fig.~\ref{fig:pics_FIG_SD}a and b we compare the analytical solution Eq.(\ref{gammasol}) to numerical simulations of the process and find a perfect agreement. More intermittent  processes, i.e.~smaller $\alpha$ lead to broader (but still exponentially decaying) stationary distributions, since longer diffusive excursions become more likely. The stationary standard deviation, $\sigma_{st}= \sqrt{D\langle  \tau \rangle (1+\alpha)/\alpha}$ only corresponds to the typical distance diffused by the particle $x_{typ}=\sqrt{2D\langle \tau\rangle }$ for $\alpha=1$. 
As already mentioned a further important example is the case of drawing the resetting times from a
Weibull distribution with density $\psi(\tau; \alpha, \gamma)= \gamma \alpha(\gamma t)^{\alpha-1} e^{{-\gamma t}^\alpha}$, see Fig.~\ref{fig:pics_FIG_SD}c and d. We can again calculate the stationary distribution analytically 
and observe the same qualitative behavior as for the Gamma distribution.
The interesting case of regular resetting at $\langle \tau \rangle$ can be obtained in the limit of $\alpha \to \infty$ of either the Gamma or the Weibull waiting time distribution (while keeping $\langle \tau \rangle$ fixed), or indeed as a limit of any waiting time distribution that approaches $\psi(\tau) = \delta(\tau - \langle \tau \rangle)$ as an appropriate limit. In this important case, where resets are precisely timed, we can express the stationary distribution concisely as 
\begin{equation}\label{det}
p^{\mathrm{(det)}}_\mathrm{st}(\tilde{x}) = \tfrac{1}{\sqrt{D \langle\tau\rangle}} \left[ \tfrac{1}{\sqrt{\pi}} e^{-\tilde{x}^2}- |\tilde{x}| (1-\text{erf}(|\tilde{x}|) \right],
\end{equation}
where $\text{erf}$ denotes the error function and $x = 2\,\tilde{x}\sqrt{D\langle\tau\rangle}$.

Finally, we consider the solution of Eq.(\ref{Laplacesol}) in the limit of small $s$. Using the moment expansion ${\hat \psi}(s)=1-s\langle \tau \rangle+\mathcal{O}(s^{2})$, we obtain from Eq.(\ref{Laplacesol}), e.g. for the diffusion case, after Laplace inversion $p_{st}(x)= (2x_{typ})^{{-1}}\exp{-\frac{|x|}{x_{typ}}}$. Observe that  for resetting governed by
a scale-free waiting time distribution with asymptotic behavior $\psi(\tau)\sim \tau^{{1-\delta}}$ with $0<\delta<1$, the intensity of events decays as $\nu(t)\sim t^{{\delta-1}}$.
Therefore, for long times the process is dominated by one large diffusive excursion and does not exhibit a NESS.

Evidently, the resetting protocol also has a strong influence on the search efficiency to find a target located at distance $x_0$ from
the origin. For the case of constant rate resetting, an $x_0$-dependent optimal rate has been obtained in \cite{Evans}. However, in many
situations the search strategy is constrained by the need to return on average after a characteristic time $\langle\tau\rangle$.
For example, many foraging animals have to return home regularly to rest because of exhaustion of energy or to feed their offspring. Our previous results indicate that the efficiency 
can also
be optimized by adapting the functional form of the distribution. 
In Fig.~\ref{fig4}a we show for Gamma-distributed resetting
times with a fixed mean resetting time $\langle \tau \rangle $ that the mean first passage time indeed exhibits a $x_0$-dependent minimum corresponding to an optimal $\alpha_{opt}(x_0)$. The existence of such a minimum can be intuitively understood by inspecting the limits of the resetting protocol. For $\alpha\to\infty$, resets occur deterministically after time $\langle  \tau \rangle $ and targets at $x_0\gg x_{typ}$ are almost surely never encountered. On the other hand for 
$\alpha\to 0$  the reset protocol is extremely intermittent and the process is typically dominated by a single large excursion. 
In this case the process resembles diffusion without resetting, with a diverging mean first passage time. To determine $\alpha_{opt}(x_0)$ we 
make the reasonable assumption that for $x_0\ge x_{typ}$ the standard deviation of the steady state for $\alpha_{opt}(x_0)$ should be proportional to $x_0$, i.e.~$\sigma_{st}[\alpha_{opt}(x_0)]/x_0=C$.
We then obtain
\begin{equation}\label{alphaopt}
\alpha_{opt}(x_0)=\left(C^2 x_0^2/(D\langle \tau \rangle )-1\right)^{-1}\, ,
\end{equation}
where the $C$ has to be determined numerically once and for Gamma-distributed resetting times is $C\!\approx\!0.83$.
In Fig.~\ref{fig4} we compare our analytical prediction Eq.~(\ref{alphaopt})
with the numerically determined values of $\alpha_{opt}(x_0)$ and find a very good agreement. For small $x_0<x_{typ}$ Eq.~(\ref{alphaopt}) is no longer valid and it is very hard to identify $\alpha_{opt}$ because the stationary distribution becomes increasingly insensitive to variation
of $\alpha$ for $\alpha>1$, see Fig.~\ref{fig4}a. These results indicate that an intermittent return strategy should be adopted when targets remote from the home may exist.


In conclusion, our study reveals that the NESS as well as the search efficiency depend on the precise form of the resetting distribution. 
Interestingly, for Continuous Time Random Walks (CTRWs) - a related class of models, where the times between successive jumps are drawn from a waiting time distribution - such a dependence of the long time behavior on the precise form of the waiting time distribution is not observed.
Here the knowledge of the characteristic waiting time is sufficient to characterize the steady state distribution of the process in a confining potential \cite{Sokolov}.
Finally, we note that, for general resetting times, the steady state of the process cannot be simply obtained by averaging the propagator over the waiting time 
distribution. Instead, one has to average over the distribution of times that have elapsed since the last reset event, the so-called backward occurrence time distribution. This distribution is only identical to the waiting time distribution for memoryless processes, i.e.~exponentially distributed waiting times. For other waiting time distributions the corresponding backward occurrence time distribution can in general not be obtained in a closed form. 



\end{document}